# COMPARISON BETWEEN AUTOSAR PLATFORMS WITH FUNCTIONAL SAFETY FOR AUTOMOTIVE SOFTWARE ARCHITECTURES

*Youssef EL KHARAZ[1], Saad MOTAHHIR[2], Abdelaziz EL GHZIZAL[3]*

[1]*author's information: EST USMBA*
*Fez, Morocco, Technopolis Rabat shore B5 Sala El Jadida*
*+212658439570, youssef.elkharaz@usmba.ac.ma*
*……………………..*

In the next Vehicle generations, connected and highly developed driving cars will have an important impact on the networking architecture and the interconnection between ECUs(Electronic Control Unit). The automotive industry begins to develop new and efficient strategies to improve the performance of the global system. AUTOSAR organization as part of this industry tries to present plenary solutions especially software architectures for new technologies in this field. Thus, in this paper, we present the aspects of new E/E architectures with upcoming technologies. We discuss a new solution presented by AUTOSAR organization to implement new software requirements for next generation cars. This solution aims to provide a safe environment for the features that require complex data processing and to communicate with AUTOSAR and non AUTOSAR Platforms. We summarize a comprehensive comparison between AUTOSAR adaptive and AUTOSAR classic in terms of functionality and application area. We provide functional Safety preliminaries for the global E/E architectures.

## 1. Introduction

The Recent cars like connected and autonomous vehicles becoming a state of art in automotive industry. That lead directly to increase the percentage of electronic components(Kassakian and Perreault, 2001) and embedded systems within an automobile. Moreover, automobiles are inherently safety critical. The use of ECUs in modern day cars are growing exponentially, each for a specific functionality. Present technologies such as infotainment(Choi et al., 2019), Car-to-X technology(Schafer and Klein, 2013), Autonomous cars(Hussain and Zeadally, 2019), etc require high level computing power. This means that the architecture network shall process and transport large size of data in a short time.

To establish these new technologies on the software Architecture besides the existing requirements including safety and security, a dynamic software environment is needed. In this context, the AUTOSAR organization formed by OEMs and suppliers recognized that these new requirements couldn't be implemented by the existing software architectures(Furst and Bechter, 2016). In general, AUTOSAR provides solutions for different software requirements throughout layered architecture that separates the software from microcontroller where the reuse of software components for other applications is workable(Bunzel, 2011), as depicted in Fig1. The AUTOSAR consortium provide two standards called Classic Platform and Adaptive Platform that are used for different goals and requirements. The classic platform is already established in the industry and intended to fulfill stringent real-time requirements with cost-optimized processors. In other hand, AUTOSAR adaptive Platform was introduced recently to handle applications with outstanding performance requirements such as highly automated driving. The combination of these trends will pave the way for the upcoming E/E architectures.

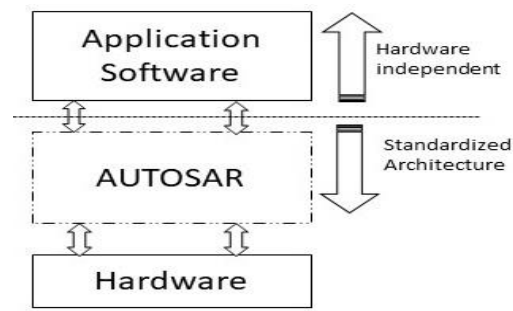

*Figure 1. Application Software separated from the hardware.*

Throughout this paper, we present the Key aspects of heterogeneous platforms, and we highlight the AUTOSAR Adaptive platform standard with its last release. We study the main differences between classic AUTOSAR and Adaptive AUTOSAR. We perform a scenario about the communication between Classic and Adaptive AUTOSAR. We provide the functional safety preliminaries to establish safely the automotive applications.

## 2. Overview of the future E/E architectures

### 2.1. Combination of different software platforms

Nowadays, networking architectures of new cars can be found in different domains, such as connectivity, infotainment, electrification, etc. Each domain has a specific ECU with a specific methodology of development. In addition of infotainment ECUs which are using Linux or others operating systems, and AUTOSAR Classic Platform is the standard for deeply embedded ECUs, another type of ECUs must arise with different characteristics that has to go along and interconnected with existing E/E architectures to respond the demands from the new embedded applications.

### 2.2. Service Oriented communication

The Communication between ECUs throughout signals using traditional protocols like CAN, LIN, FlexRay and MOST fit very well for transmission data within an automobile. Meanwhile advance technologies like infotainment and Car-to-X demand higher bandwidth which are not highly supported by these protocols because of limited characteristics(Furst and Bechter, 2016).

The service-oriented communication(Gopu et al., 2016) is a flexible and efficient way to interconnect systems and their subscribers based on applications which provide services on the communication network. The future E/E architecture will be strictly based on the combination of the service-oriented paradigm and the existing traditional communication. For that AUTOSAR as a consortium formed by OEMs and their suppliers are in charge to standardize a new platform which is called AUTOSAR Adaptive platform using existing standards.

## 3. The AUTOSAR Adaptive platform

The AUTOSAR Adaptive platform is a standardized architecture for high-performance ECUs to build safety systems such as highly driving and autonomous systems. Figure 2 depicts the global architecture of this platform.

Starting from the release 1.0.0 until the last release R20-11, several concepts affecting the Adaptive Platform have been introduced thereby adding new functionality to the platform, one of the core features of this adaptive platform is called AUTOSAR Runtime for Adaptive Applications (ARA). ARA gives users all the interfaces and infrastructure needed to communicate and execute adaptive applications into the system and allows data exchange between ECUs regardless of their internal architectures. In addition, this runtime offers direct access to the operating system functions known as the "Minimum Real time system Profile" (PSE51).

The module operating system interface based on a subset of POSIX (Atlidakis et al., 2016) is responsible for run-time resource management such as signals, timer and thread handling for all adaptive applications and functional clusters that establish the platform.

In AUTOSAR Adaptive platform, Applications are not totally bounded by a static scheduling and memory management but are free to allocate memory on their current need and break down their tasks thanks to object-oriented programming.

The Execution Manager module is an element of the architecture responsible for startup and stopping the AUTOSAR Adaptive Applications, and responsible for providing the necessary resources during the execution period of the applications. To ensure the communication between local applications and applications on other ECUs including the interaction with the Adaptive platform services, a middleware protocols must be defined. The most noticeable changes in the use of AUTOSAR Adaptive are the universal use of Ethernet based communication systems. For the Release R20-11 new technology added to support the Ethernet protocol is related to 10BASE-T1S(Huszak and Morita, 2019)which is specified by

IEEE802.3cg. This new feature allows easy integration devices into automotive Ethernet using the multi-drop configuration. Furthermore, it is localized on layers 1 and 2 of the OSI model is to be supported by Classic Platform as well as Adaptive.

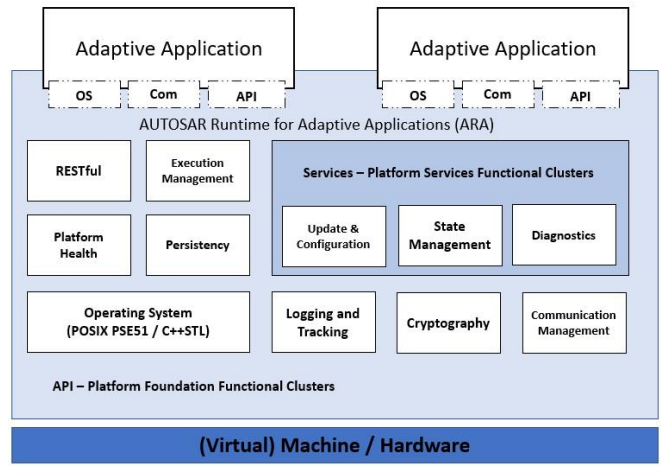

*Figure 2. Adaptive AUTOSAR platform*

AUTOSAR organization has an extensive release plan for adaptive AUTOSAR. The latest Release date was in November 2020. The main focus for this release is to enhance the security and communication (10BASE-T1S, ara Communication Groups) by adding new functionality to the platform. Additionally, some concepts target the Classic and Adaptive Platform and reinforcing the interaction between the two platforms.

## 4. AUTOSAR Foundation

ECUs of future cars consists of different architecture platforms to offer required functionalities such as highly automated driving, connectivity, chassis, and infotainment, note that each system might be classified as a safety-critical or no safety-critical part. It's necessary a middleware communication between platforms to achieve complete functionality of the global system. For that, the AUTOSAR organization defined a separate standard called AUTOSAR Foundation. The main goal of the Foundation standard("Foundation Release Overview," n.d.) is to enforce interoperability between the AUTOSAR platforms, and to ensure compatibility between:

- Classic- and Adaptive Platform.
- Non-AUTOSAR platforms to AUTOSAR platforms.

## 5. Classic AUTOSAR Vs Adaptive AUTOSAR

Classic AUTOSAR was the first achievement of a standardized software architecture for ECUs, then Adaptive AUTOSAR has just supplied a new architecture to meet new OEM requirements. In the classic platform, the application layer handles the communication between software components through the runtime environment (RTE) with the help of OS. The RTE acts as an abstraction layer in between ECUs and establishes inter-ECU communication via a specific communication network. In the Adaptive platform, the applications utilize the "AUTOSAR Runtime for Adaptive Applications," also known as ARA.

This runtime environment gives users standardized interfaces to efficiently integrate different applications into the system. ARA offers mechanisms for ECU-internal and inter-network communications as well as access to basic services such as diagnostics and network management. The adaptive AUTOSAR applications are formed in software components that communicate via services. These services may be requested or provided. In addition, the application programmer can directly access a subset of operating system functions. In terms of communication, the AUTOSAR Adaptive defines a new feature called ara::com. ara::com is a standard C++ API based on SOA more specifically based on SOME/IP. In parallel rough, AUTOSAR classic cover the communication between software components using RTE signals.

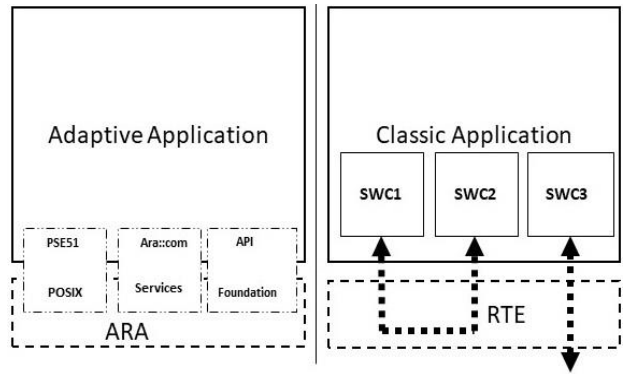

*Figure 3. Difference between Applications within Adaptive AUTOSAR and Classic AUTOSAR.*

In the classic platform, the application layer handles the communication between software components through the runtime environment (RTE) with the help of OS. The RTE acts as an abstraction layer in between ECUs and establishes inter-ECU communication via a specific communication network. In the Adaptive platform, the applications utilize the "AUTOSAR Runtime for Adaptive Applications," also known as ARA.

This runtime environment gives users standardized interfaces to efficiently integrate different applications into the system. ARA offers mechanisms for ECU-internal and inter-network communications as well as access to basic services such as diagnostics and network management. The adaptive AUTOSAR applications are formed in software components that communicate via services. These services may be requested or provided. In addition, the application programmer can directly access a subset of operating system functions. In terms of communication, the AUTOSAR Adaptive defines a new feature called ara::com. ara::com is a standard C++ API based on SOA more specifically based on SOME/IP. In parallel rough, AUTOSAR classic cover the communication between software components using RTE signals.

The operating system presents some of the key differences between the adaptive AUTOSAR platform and the classic AUTOSAR platform. In adaptive AUTOSAR applications are no longer tied to very strict and static scheduling and memory management but are certainly free to create and destroy tasks and allocate memory as needed, including the use of C++ as programming language in contrast to C in classic platform.

```
Void Function_Runnable(){

Uint8 Data_Input;
Uint8 Data_Output;

RTE_ReadSignal_Input(&Data_Input);

/* brief Implementation*/

RTE_WriteSignal_Output(&Data_Output);
}
```

```
Class FuncionServiceProxy{
public:
        /* brief implementation of FuncionServiceProxy class*/
Explicit FuncionServiceProxy(HandleType &handl);

        /* brief public member for the BrakeEvent*/
events::BrakeEvent Brake vent;

        /* brief Public Field for UpdateRate*/
fields::UpdateRate UpdateRate;

        /* brief public member for the Adjust method*/
methods::Calibrate calibrate;

        /* brief public member for the Adjust method*/
methods::Adjust Adjust;

};
```

*Figure 4. An example illustrates the code implementation of RTE and ARA::COM.*

An important point in AUTOSAR platform is update and Configuration Management. The Adaptive platform now offers the option of removing, updating, or adding individual applications, while the Classic platform can only replace the entire ECU code during an update.

Another property of the Adaptive platform is its transition to an exclusively service-oriented architecture paradigm, which offers greater flexibility in system design. Applications provide their functionality as a service via the Adaptive platform, and they can use services that are offered. In other hand the focus of the Classic platform is primarily on signal-oriented communication. Nonetheless, it is also possible to use AUTOSAR Classic in a service- based way for communication between multiple ECUs. In practice, the main properties of the AUTOSAR Adaptive and Classic platforms complement one another. It may therefore be assumed that ECUs based on both standards will be used in future vehicles resulting in a heterogeneous architecture.

**Table 1:** Comparison between AUTOSAR adaptive and AUTOSAR classic.

|  | **AUTOSAR Classic** | **AUTOSAR Adaptive** |
|---|---|---|
| **Application Interface** | Use RTE with the help of OSEK | Use ARA |
| **Operating system** | OSEK | POSIK |
| **Programming language** | C | C++ |
| **Remove/update application** | Remove or update the entire ECU | Remove, add, or update individual application |
| **Communication protocols** | Signal based communication network bus (CAN, LIN, etc) | Service oriented communication based on ethernet over (SOME/IP) |
| **Utilization** | Implement deeply embedded systems | Implement high performance applications like high-automated driving |
| **Functional safety** | Up to ASIL D | Up to ASIL D |

## 6. Communication between AUTOSAR Platforms

How to communicate between Adaptive and Classic ECUs is a mandatory question. In such a scenario, the ECUs which are interconnected over Ethernet use service-oriented communication over SOME/IP. In this example, the AUTOSAR Classic ECU1 is connected to multiple bus systems to which other ECUs are connected (Fig.5).

ECU1 operates as a gateway in this configuration and it's responsible for transferring the message signals from the bus side into a service so that they can be accessed directly by the AUTOSAR Adaptive platform. The communications layout is a fixed component of the design of AUTOSAR ECUs, whether it is a Classic or Adaptive platform. Because the configuration format is different for the two platforms, it is necessary to map the service configuration in the form of a conversion. The situation is somewhat more multifaceted for communicating with an AUTOSAR Classic ECU whose operation is exclusively signal based. In this scenario, the ECU1 is designed as a signal gateway, and it converts message signals directly into UDP frames("Specification of UDP Network Management," n.d.) on Ethernet. The AUTOSAR Adaptive ECU now converts signals from the UDP frame to a service that is available within ECU2.

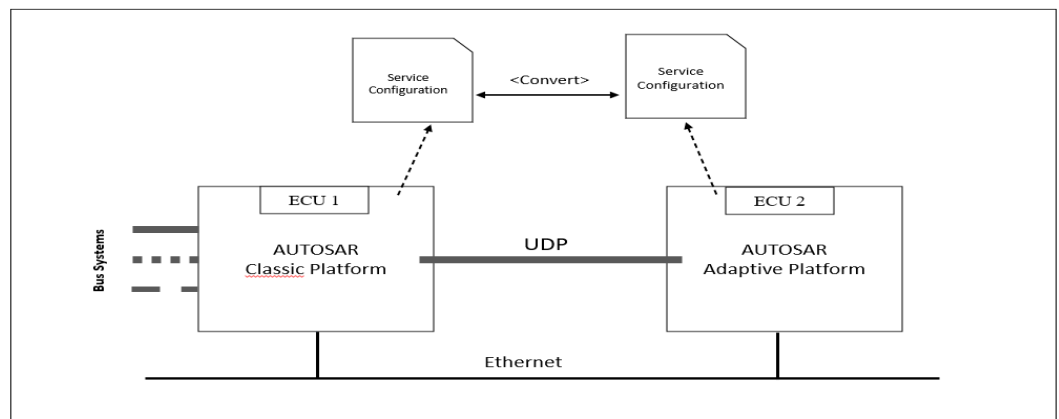

*Figure 5. An example illustrates the code implementation of RTE and ARA::COM.*

## 7. Safety preliminaries for E/E Architectures

### 7.1. Functional Safety overview

In the Automotive Industry, we prefer to define Safety as freedom from hazardous situations that can cause damage, physical or material for humans in the car and on the road.

During the process of realizing the Automotive applications, guaranteeing safety is always a prerequisite. According to the definition in ISO 26262, functional safety Means no unpredictable faults due to hazards Risks by the malfunctioning behaviour of E/E Architecture("ISO - Standards," n.d.).The standard ISO 26262 has two versions, the first which has been introduced in 2011, and the recent one in December 2018("ISO 26262-1:2018(en), Road vehicles — Functional safety — Part 1: Vocabulary," n.d.).

### 7.2. ASIL Determination.

In ISO 26262, Automotive Safety Integrity Level (ASIL) is a risk classification scheme that helps the developers to meet the functional safety requirements for safety automotive applications. The ASIL is based on Hazard Analysis and Risk Assessment through estimating the severity, exposure, and controllability of each Hardware/software component behaviour. ISO 26262 identified four ASILs Levels A, B, C and D plus Quality Management (QM) which is not considered as an ASIL, while ASIL D represent the highest degree of automotive hazard and ASIL A represent the lowest degree.

To determine an ASIL we need a combination of severity, exposure, controllability. Exposure Indicates the probability of causing harm, severity Indicates the extent of harm, and controllability Indicates the ability to avoid the harm through the timely reactions of the drivers(Xie et al., 2020).Table-II illustrate the ASIL Determination based on severity, exposure, and controllability.

**Table 2:** ASIL determination in ISO 26262

| Severity | Exposure | Controllability | | |
|---|---|---|---|---|
| | | C1 | C2 | C3 |
| S1 | E1 | QM | QM | QM |
| | E2 | QM | QM | QM |
| | E3 | QM | QM | A |
| | E4 | QM | A | B |
| S2 | E1 | QM | QM | QM |
| | E2 | QM | QM | A |
| | E3 | QM | A | B |
| | E4 | A | B | C |
| S3 | E1 | QM | QM | A/QM |
| | E2 | QM | A | B |
| | E3 | A | B | C |
| | E4 | B | C | D |

## 8. CONCLUSION

Currently, Autonomous, and connected cars technology are developing and presenting more challenges especially in terms of safety. In the software part, AUTOSAR as an organization provides solutions that fit very well for these challenges.

Adaptive AUTOSAR is not designed to replace Classic AUTOSAR in terms of functionality, but we need them both to coexist and cooperate to meet the needs and challenges of the automotive industry. To determine which one to use in a specific application, we need to start by answering the following questions: How much computing power do we need? What are our time requirements? And how dynamic is our application?